# Reel Stock Analysis for an Integrated Paper Packaging Company

Constantine Goulimis, Gastón Simone

Greycon Ltd., 7 Calico House, Plantation Wharf, London SW11 3TN.
{cng✉,gs}@greycon.com

**Abstract**

The production of corrugated paper boxes accounts for roughly one third of the world's total paper production and, as a result of both COVID-19 and the rise of e-commerce, is a growing market. We provide a fresh approach to determining near-optimal stock policies for integrated paper companies. The new approach shows that existing policies can be improved by a significant margin. In a case study we saw a reduction in total waste by 9%, with a simultaneous decrease in logistics costs.

## 1 Introduction

Corrugated box production accounts for approximately 1/3 of global paper / board production and demand is rising, driven by the increase in deliveries and the recyclability of the material. Corrugated boxes are produced by *corrugating plants*, which in turn consume paper produced by *paper mills*. An *integrated paper packaging company* will have both paper mills and corrugating plants. Typically, the latter outnumber the former by a factor of 5 or 10.

Production of corrugated boxes took off in the 1950's with the introduction of the *corrugator*, a machine that produces the multi-layer rectangular sheets that are then cut and folded into a box (see https://en.wikipedia.org/wiki/Corrugated_fiberboard). A modern corrugator costs tens of millions of euros / dollars and would be 2.5 m wide. It is fed by paper produced on a paper machine. A typical modern paper machine will cost hundreds of millions of euros / dollars and would be 5.0 – 9.0 m wide. Since boxes tend to be no wider than ~1.0 m, the mismatch of widths gives rise to a linked cutting stock problem (*CSP*).

Because the corrugating plants operate an order-to-delivery cycle time of days if not hours, paper inventory is held at the corrugating plants. This inventory, replenished from the paper mill(s), is used on the corrugators to produce the sheets.

A typical corrugated material will contain three layers of paper, each of different grade and thickness (for brevity we talk of grade, when in fact each grade can come in a variety of standard thicknesses):

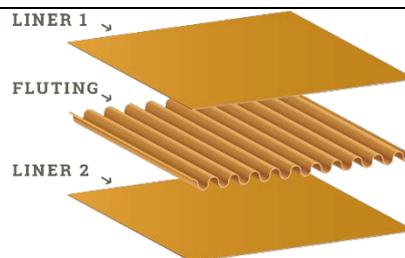

**Figure 1.** The three layers of corrugated material. A corrugator will consume reels of paper for each of the layers and produce rectangular sheets of the required dimensions.

Different box types will share some paper grades, particularly for the fluting and less so for the top/bottom (but there is some sharing even there). For the fluting layer, in terms of surface area, 1 m² of box will require ~1.2 m² of the middle layer; the precise number is determined by the wavelength of this layer. In effect, there is a bill of materials (*BoM*) with an entry for each layer of a box sheet. There are so many possible variations in box characteristics (dimensions, BoM, cutouts, folding creases, printing) that in practice boxes are exclusively made-to-order.

The above is a highly simplified description of the process. There are many real-world constraints that impact the analysis; we list some of them later on.

At a corrugator there is an opportunity to combine orders of different dimensions, but with identical BoM. Because the corrugated board cannot be wound into reels (it has to remain flat), it is cut into sheets in-line; most corrugators are *double-knife*, i.e. can only handle two distinct sheet lengths at any time. This gives rise to a well-understood and solved semi-continuous variant of the CSP (the first published reference we can find is in 1964 [1]; there are many companies supplying software for solving this problem for the corrugating industry; consequently, there are few academic publications). Figure 2 depicts a small, but otherwise typical, corrugator instance:

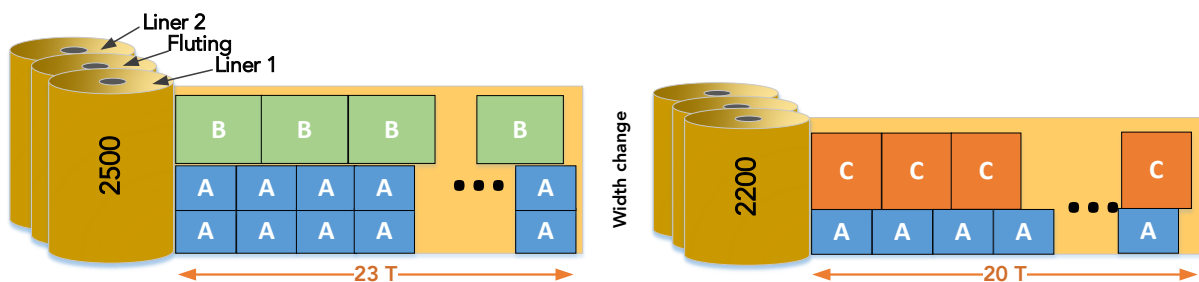

**Figure 2. Schematic of a corrugator CSP: Three sheet sizes (A, B & C), sharing a common BoM, are produced on a single corrugator using two paper reel widths (2500 mm & 2200 mm). After producing the first pattern (2 × A + B) from paper 2500 mm wide, the paper width is changed to 2200 mm to produce the second pattern (A + C). A double-knife corrugator, such as the one depicted here, only allows two different sheet lengths to be produced in any one pattern (so, the combination A+B+C would be infeasible).**

Other things being equal, it is preferable to run a corrugator at its maximum width. However, this may generate too much waste (consider e.g. producing the second combination with orders A & C out of the bigger reel width) and this is what induces multiple reel widths for the inventory. Industry practice is to hold 8-12 reel widths (for each paper grade / grammage combination) in inventory at the corrugating plant. We call the list of reel widths stocked at the corrugating plants ({2200, 2500} in this example) the *inventory policy*.

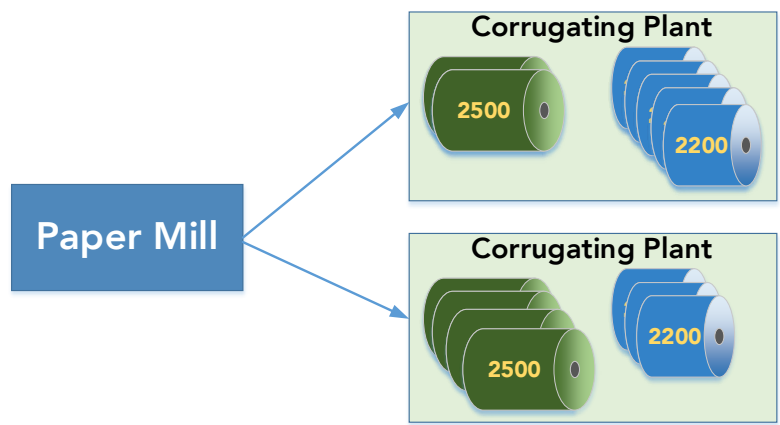

**Figure 3. A paper mill supplying two corrugating plants, with inventory policy {2200,2500}**

Back at the paper mill, the demand from the corrugating plants (e.g. for the 2500 mm and 2200 mm reels) leads to another cutting stock problem. It is possible, indeed likely, that there is a trade-off between the reel widths that suit the corrugator(s) and those that suit the paper machine(s). So, the fundamental question addressed by this paper is to explore this trade-off and determine a near-optimal policy.

## 2   Previous Work

There appears to be very little published work on this problem. The first related reference we could find is from 1983 in [2]:

> "The maximum width roll of liner and medium that can be processed by a corrugator at a box plant is a function of the design of the corrugator. In addition to the maximum size roll, most plants stock a number of narrower width rolls. This decision obviously leads to an increase in box plant inventory and a reduction in corrugator productivity. The primary motivation for stocking multiple roll widths is to reduce side trim generated at the corrugator. Because side trim is such an easily measured factor, it generally dominates in the stock size selection process. This can easily lead to a proliferation of stocking sizes. The costs associated with increased inventory and reduced corrugator utilization is more difficult to isolate and measure. This inhibits making an economic tradeoff to determine the number of sizes to stock."

However, this does not consider the trim efficiency at the paper machine(s) and is therefore closer to the classical assortment problem [3]. Several commercial suppliers that offer corrugator CSP software also offer "roll stock analysis", which aims to recommend inventory policies based on historical corrugator demand. However this "analysis" is done, it only looks at the corrugator efficiency, ignoring the impact on the paper machines. The sub-optimality of the resulting solution is made more stark when we consider that the waste generated at the paper machines is often significantly higher than that at the corrugators (more about this later).

In their otherwise comprehensive look at scheduling issues in the paper industry of 2001, [4] do not discuss our problem at all. Similarly, it fails to appear in the 2009 survey [5] and follow-on 2019 survey [6] of supply chain problems in the paper industry.

We speculate that the reason for the paucity of relevant research is the computational intractability of the underlying problem, perhaps coupled with the need to link statistical aspects with combinatorial optimisation ones.

## 3   Practical Aspects

We list below some of the aspects and constraints that may be necessary in any real-world implementation:

### 3.1   External Demand

It is possible that some of the paper grades used for corrugating boxes are supplied by external suppliers. These do not need to be taken into account at the paper machine CSP. Conversely, it is possible that the paper machines also produce paper for external demand; such demand does not directly impact the corrugating plants, but affects the CSP instances at the paper machines.

The presence of external supply of / demand for paper complicates the notion of percentage waste. This is best illustrated with an example: A paper mill produces 1,000 T, with 3% waste (30 T). Of the 970 T net production, 600 T goes to the open market and the corrugators consume the remaining 370 T. The corrugators also buy, from external suppliers, 130 T. Their input material consumption is therefore 500 T, and they incur 2% waste (10 T) on this. It is hard to defend any specific total percentage waste calculation in the circumstances. We use the convention of adding up the total waste (30 T + 10 T = 40 T) and expressing this as percentage of the total paper machine production, i.e. 4.0% in this case.

### 3.2 Triple-knife Corrugators

Most of the world's corrugators are *double-knife*, meaning they can cut up to two different sheet lengths at a time, as shown in Figure 2. For those, industry practice is to keep 8-12 reel widths in inventory. However, *triple-knife* corrugators exist; these can combine three order lengths at a time. This additional flexibility reduces the number of reel widths in a near-optimal policy down to one or two.

### 3.3 Different Materials / Box Types

Most corrugated boxes have three layers. However, 5-layer (*double-wall*) boxes are used for heavier objects. Also, some packaging scenarios call for a 2-layer material (*single-facer*).

### 3.4 Downgrades / Upgrades

Material substitution is common practice in the industry. For instance, you could substitute a 200 g/m² paper for one with 180 g/m², therefore giving the end customer a slightly higher specification. The additional cost may be offset by other factors such as reduced trim waste or inventory / logistics costs.

### 3.5 Demand Assignment to Corrugators

When an integrated paper company receives an order for a box, it typically has to decide which corrugating plant to assign this to (box orders are almost never split across plants). This is typically decided by transportation cost and production capability considerations. Once the corrugating plant has been determined, there is a further decision as to which specific corrugator will produce this order (orders are almost never split across corrugators).

### 3.6 Paper Machine Production Cycles

Almost all paper machines operate a production cycle, e.g. grade Testliner, basis weight 180 g/m² is produced every second week, followed by the same grade, basis weight 200 g/m², etc. Although there is evidence that some of these cycles are sub-optimal, it is very difficult to change them, as well as being a difficult-to-solve (non-convex) optimisation problem.

### 3.7 Multi-paper-machine Scenarios

In the general case, there will be multiple paper machines in multiple paper mills. As a result, there is a sourcing decision to be taken. For example, two different paper machines, of differing widths, potentially located in different mills (thus inducing differences in transportation costs) could source the same paper grade.

### 3.8 Policy Uniformity

An integrated paper company will have many corrugating plants, each potentially with multiple corrugators. A *uniform* policy means that at each plant, the items that are inventoried are a subset of the overall policy. This is dictated by corrugator constraints. For instance, one plant may have a large corrugator, which is 2700 mm wide; another plant's maximum corrugator width may be 2500 mm. There is no point in the second plant holding inventory of items wider than 2500 mm. Even with a uniform policy, the quantities of each of its elements will vary by plant (as shown in Figure 2).

However non-uniform policies are also possible: each of the plants may have a policy of specific cardinality, but the policies can vary across corrugating plants. For example, one corrugating plant may specialise in boxes for agriculture, another plant may specialise in boxes for e-commerce. The dimensional characteristics of these could be different and there is no need for the inventory to have any common reel widths.

We present here results for uniform policies alone, but the approach naturally extends to non-uniform ones as well.

At a single corrugating plant one might have thought that in general different paper grades need different policies. However, the fact that boxes are made up of several grades creates a coupling constraint: the three input reels at any given time (for each of the box's layers) must be of the same width. The middle layer in particular (*fluting*) is common across many different types of boxes. In practice, a common policy across paper grades at a single plant is the industry norm.

### 3.9 Paper Cost Variability

Paper is sold by weight, reflecting the underlying material, energy and capital costs. Therefore one talks of cost per tonne for the material. However, different grades / grammages may have different production costs per tonne. This is important in our context because we want to determine the cost of a policy: if the production costs are essentially identical across grades / grammages, then we can just add the weight of the waste. On the other hand, if the production costs vary significantly, we would want to create a financial objective function reflecting the cost of the waste.

### 3.10 Transportation Cost

When waste is produced at the paper mill, then it is typically recycled on-site and does not incur any transportation cost. However, waste generated at the corrugating plant has to be transported back to a paper mill for recycling. Depending on the distance and the transportation modes, there might be differences between corrugating plants for these costs.

### 3.11 Static vs. Dynamic Policies

One of the fundamental difficulties in all types of inventory policy is how to adjust them to reflect demand that varies over time. As the underlying problem is combinatorial, the risk arises that an optimal policy for a particular time period is completely different from an optimal policy for another period, even if nothing else changes.

Whilst this is certainly true, we have to face the reality that integrated paper companies lack optimisation-based tools to evaluate, at least periodically, their policy. As a result, companies are often locked into policies that are inherited and sub-optimal.

## 4 Approach

### 4.1 Model & Objective Function

The fundamental building block of our approach is the *policy* denoted by *P*. This consists of a set of reel widths that will be kept in inventory. For example, one such policy consists of {1350,

1770, 1830, 1960, 2040, 2160, 2260, 2435, 2500}. We call | P | the *cardinality* of P; in this example it is 9. The determination of a cost for P involves three steps:

1. Determining the waste at the corrugators for policy P
2. Determining the waste at the paper machine, using the results from the solutions that were obtained at the previous step
3. Combining the two waste measures

We describe each of these steps below.

### Corrugator Waste

Given a policy P and a particular CSP instance for the corrugator, we can determine $w_{cor}^P$, the minimum waste when using P to supply the instance by solving the corresponding CSP. For example, for the instance:

| Item № | Width (mm) | Length (mm) | Quantity (sheets) |
|---|---|---|---|
| 1 | 1150 | 1210 | 6500 |
| 2 | 1030 | 980 | 7000 |
| 3 | 660 | 920 | 26000 |
| 4 | 610 | 750 | 10000 |
| 5 | 580 | 500 | 18000 |

and policy P = {2200, 2500}, one waste-optimal solution is:

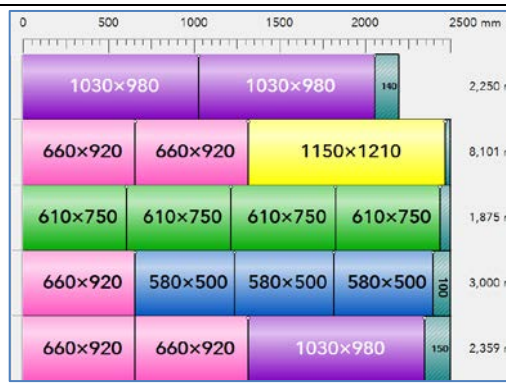

**Figure 4. Solution to corrugator CSP instance. In this case $w_{cor}^P$ = 239 kg (3.071%) and we require 891 kg of 2200 mm and 6901 kg of 2500 mm. If the BoM for this box type were Liner_A-Fluting-Liner_B in a weight ratio of 40-20-40, then the solution consumes 40% × 891 kg = 356 kg of Liner_A / width 2200 mm, etc.**

So, given a policy P and a corrugator CSP instance, we can determine

1. The cost (waste) of an optimal solution
2. The quantity required for each element (grade & width) of the policy.

We observe that both of these are highly dependent on the actual policy. E.g. changing from {2200, 2500} to {2125, 2500} (a change of only 75 mm in one of the policy elements) instead generates an answer with:

1. $w_{cor}^P$ = 200 kg (2.584%)
2. We require 1912 kg of 2125 mm and 5838 kg of 2500 mm.

At the end of this iterative process we will have determined the total corrugator waste $w_{cor} = \sum_n w_{cor}^n$ and the time-phased quantity for each of the policy's elements (e.g. we require 76 T of width 1830 mm of a specific grade in week 23).

### Paper Machine Waste

In the previous step we have constructed the time-phased demand per grade at the paper machine(s). For example, for one particular grade, in one time period, we might end up with:

| Item № | Width (mm) | Quantity (kg) | Quantity (reels) |
|---|---|---|---|
| 1 | 1050 | 18009 | 16 |
| 2 | 1250 | 16080 | 12 |
| 3 | 1400 | 55530 | 37 |
| 4 | 1600 | 30873 | 18 |
| 5 | 1900 | 42773 | 21 |
| 6 | 2000 | 109344 | 51 |
| 7 | 2100 | 123817 | 55 |
| 8 | 2200 | 11792 | 5 |
| 9 | 2300 | 49313 | 20 |
| 10 | 2350 | 25192 | 10 |
| 11 | 2400 | 64319 | 25 |
| 12 | 2450 | 52528 | 20 |
| 13 | 2500 | 206360 | 77 |

On a paper machine that is 6000 mm wide, an optimal solution (allowing a tolerance of -0% / +5% on the quantity) to the corresponding CSP is:

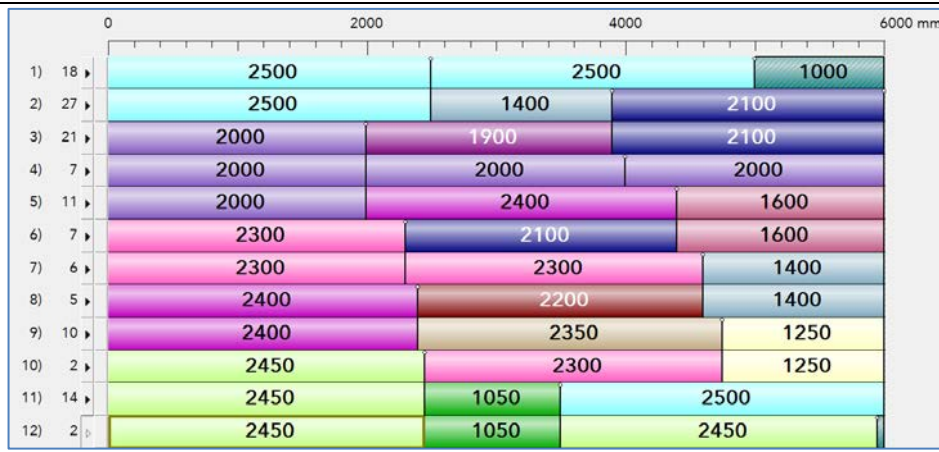

**Figure 5. Optimal solution to paper machine CSP, with one paper machine, waste $w_{pm}^P$ = 19403 kg (2.321%)**

We address the more general case, where there are multiple paper machines potentially supplying the same paper grade, by solving a multi-machine CSP. This has similarities with the corrugator CSP, when using a policy with multiple widths. Figure 6 shows the same instance, but with two paper machines, 4300 mm & 6000 mm wide:

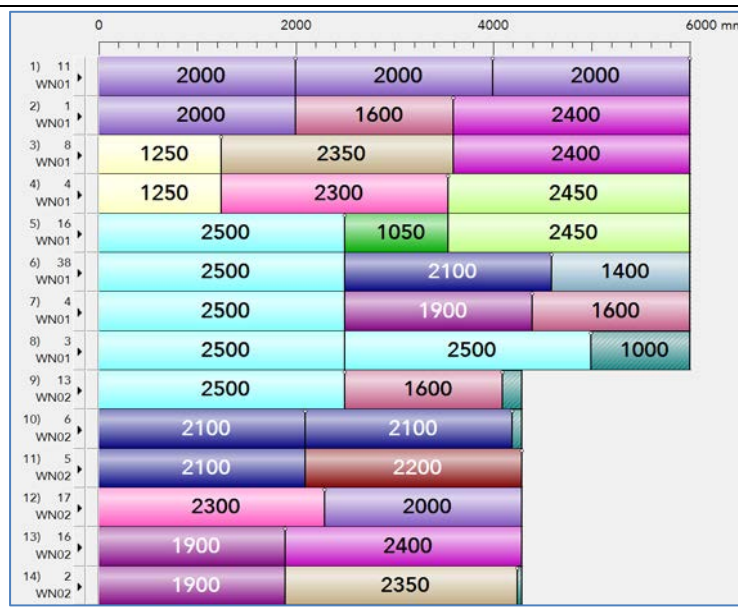

**Figure 6. Optimal solution to paper machine CSP, with two paper machines, waste $w_{pm}^P$ = 6754 kg (0.825%). Note how some sizes (e.g. the 2400 mm) are sourced from both paper machines.**

By summing the waste across all paper machine CSP instances, we will have determined, for policy P, the total waste at the paper machine, $w_{pm}$.

### Policy Cost

The overall cost of the policy is then defined to be $z = \beta w_{cor} + w_{pm}$, $\beta > 1$. The reason for $\beta > 1$ is that waste at the paper machine can be recycled on-site, whereas waste at the corrugator needs to be transported back to the paper mill(s).

### Observation I – Non-linearity

It is clearly the case that if we take two CSP instances A & B (for the same master sizes) and we merge them, then the waste of the resulting solution will be no more than the sum of the original instances:

$$w_{A \cup B} \leq w_A + w_B$$

This is because all patterns that are feasible for A and B, are also feasible for A∪B. In practice, we would expect, most of the time, a strict inequality to apply. This non-linearity stops us from assembling a "giant" CSP instance, one for each corrugator. Instead, we have to use a statistical approach, where the corrugator CSP's are as representative as possible of the real ones.

In addition, one might imagine that, because the widths of a policy *P* are common across all paper grades, we would have to solve only one instance at the paper machine. However, the required quantities are all different and this precludes us from using such a shortcut.

### Observation II – Non-dominance

Suppose that we have two policies, $P_1, P_2 : P_1 \subseteq P_2$. We can easily see that

$$w_{cor}^{P_1} \geq w_{cor}^{P_2}$$

because every solution for policy $P_1$ is also valid for $P_2$. For the corrugator, as we increase the cardinality, the optimal waste cannot increase.

However, we cannot draw a similar conclusion for the waste at the paper machine stage. A trivial example would be to compare the policies $P_1$ = {2500} and $P_2$ = {2200,2500} for a paper machine that is 5000 mm wide. Policy $P_1$ incurs no paper machine waste, whereas $P_2$ most definitely does.

It follows that we cannot draw any conclusions regarding $z = \beta w_{cor} + w_{pm}$. Somewhat counter-intuitively, an optimal policy with cardinality *n* can be better overall than one with *n+k*; this phenomenon is observed in practice.

## 4.2 Policy Search

We have outlined above a process for assigning a figure of merit, related to the waste / cost, for an inventory policy *P*. But how can we search efficiently in the space of *P*?

Determining the objective function for each *P* is time-consuming. In a mid-sized real-world study there would be several thousand corrugator CSP instances and a few hundred paper machine CSP instances. In the test-case we report below, it takes ~30 minutes on a modern computer to determine the cost of a policy.

Searching the inventory policy space is problematic, particularly as each policy evaluation is so time-consuming. We started first searching for policies with the same cardinality as an initial, baseline, policy. Can we derive some information to guide the search? One way to do this is to assign a waste to each element of the policy, so that we can distinguish between "good" and "bad" reel widths. This is straightforward for the corrugator waste. It is not so clear for the paper machine case; we use the logic illustrated in Figure 7:

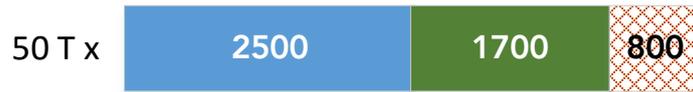

**Figure 7. Example 50 T pattern produced on a 5000 mm paper machine. The pattern wastes 800 mm. We attribute the waste proportionately to the constituent elements, so the 2500 mm is assigned 50 × (800 / 5000) × (2500 / 5000) = 4 T, etc.**

In this manner, we can distribute the total waste to each of the elements of the policy. This is a kind of sensitivity information for each reel width.

We used a tabu local search: starting from an initial policy we perturb it using a variety of "moves", some of them random, others based on the sensitivity information obtained as above. The end result carries no optimality guarantee of course.

Every analysed policy $P$ is compared against best policy found so far, $P^*$. If the new policy is worse than the best, but "close", it is worth to explore further, by applying more perturbations to it. In order to limit the number of policies we perturb, we determine whether a policy is close enough to the best, by using a threshold $\varepsilon$, initially set to $\varepsilon = 30\%$. If the new policy is within $\varepsilon$ of the best, then it is close enough and we allow perturbations. This promotes diversification. In order to favour intensification as the search progresses, we apply a decay factor $\delta = 0.05$ on the threshold as a function of the number of analysed policies $N$:

$$\frac{z(P)}{z(P^*)} \leq 1 + \varepsilon e^{-\delta N}$$

If the above inequality holds, $P$ is explored further.

Besides encouraging diversification, the search follows a "depth-first" approach, by perturbing and evaluating the best policies first. For example, if $P$ is evaluated and turns to be the new best, it will be perturbed a number of times and the resulting new policies will be placed first in a prioritised queue of policies to analyse. This logic is applied, analogously, for each evaluated policy, no matter whether it is a new best.

Although evaluating a policy is a time-consuming task, each can be executed independently of the others. So, we can examine in parallel as many policies as the available hardware allows. Taking the number of physical processors (cores) proved to be a good value for the number of policies that can be evaluated simultaneously.

Since the heuristic has no optimality guarantee and a lower bound is very difficult to obtain, we are reduced to using statistical techniques ([7], [8]) for estimating the optimal value / optimality gap. See Figure 8 for a typical distribution of the objective function in one scenario with 436 policy evaluations:

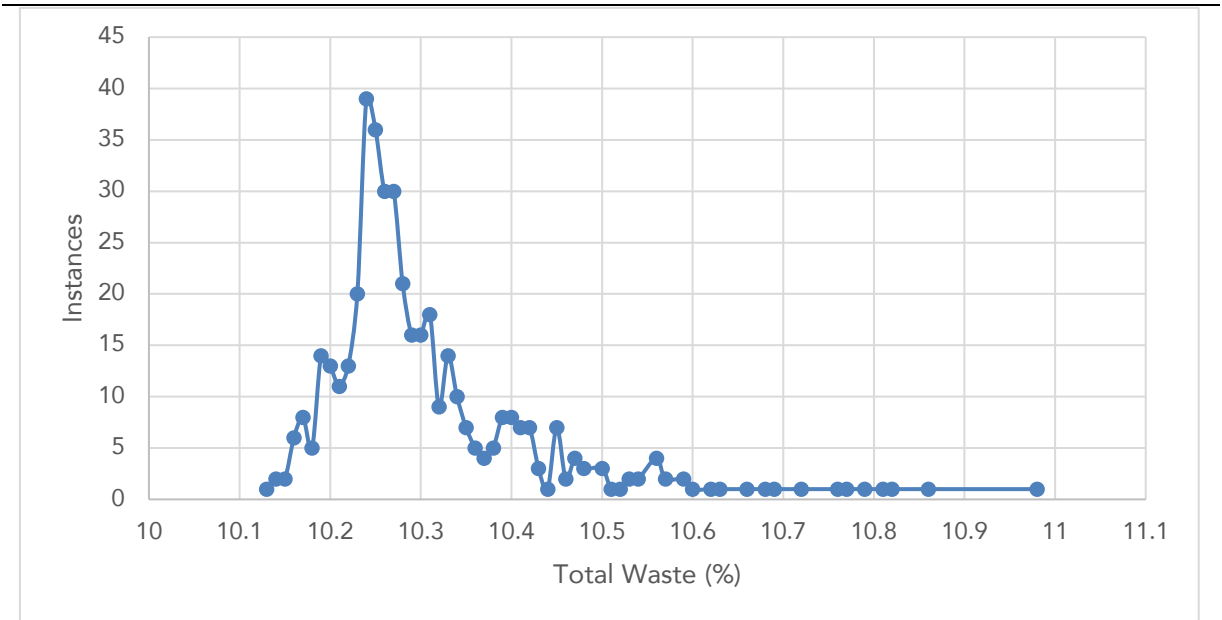

**Figure 8. Distribution of objective function (total waste in %) for 436 policies examined during the execution of the heuristic.**

We estimate the global minimum as the location parameter of a Weibull distribution fitted to the data. For this scenario, the A* estimator for the location parameter of the Weibull distribution as recommended in [9] is 10.123%, remarkably close to the minimum actually achieved of 10.131%.

# 5  Case Study

We were asked by one of our clients, with one paper mill (with a single paper machine) and three corrugating plants, to examine the following two questions:

a. Is their current inventory policy optimal?
b. If they were to replace one of the 2.5 m wide corrugators, should they buy a wider one at 2.8 m or keep to the existing size?

The data for answering these two questions consists of 5,362 corrugator instances and 124 for the paper machine.

We present results for the unrealistic case of $\beta = 1$, i.e. we treat paper machine and corrugator waste as cost-equivalent, as a means of shrouding the commercially sensitive actual results.

For question (a) we started with our client's initial policy, which had cardinality 11. The total waste for their current policy was 11.08%. The policy search identified a policy with total waste of 10.131%, a reduction of around 9%. We were also able to identify a policy with cardinality 9 (two fewer than the baseline) with even lower total waste of 9.95%.

For question (b), the empirical result was that a 2.8 m corrugator is highly advantageous without requiring an increase in the policy cardinality. With the same cardinality of 11, the total waste drops to 8.88%. This is quite surprising because the fact that the new corrugator is wider implies fewer reel widths are available for the existing ones.

# 6 Conclusions and Further Work

The approach we have presented here is promising in identifying policies that improve on the *status quo*.

However, due to the heuristic nature of the search, we are troubled by the lack of a non-statistical lower bound and this could be an interesting topic for future research. One step that would facilitate this research is improving the computational efficiency of the process – the more policies we can evaluate in a given period, the more confident we can be of the results.

# 7 Acknowledgments

The authors acknowledge the valuable comments by Sophia Drossopoulou and Daniel Carter on an earlier draft.

# 8 References

[1] S. C. Wade, "1620 corrugator trim and schedule program," IBM Corporation, Kingston, NY, 1964.

[2] R. W. Haessler, "Production Planning and Scheduling for an Integrated Container Company," in *IFAC Instrumentation and Automation in the Paper, Rubber, Plastics and Polymerisation Industries*, Antwerp, Belgium, 1983.

[3] A. I. Hinxman, "The trim-loss and assortment problems: a survey," *European Journal of Operational Research,* vol. 5, pp. 8-18, 1980.

[4] P. Keskinocak, F. Wu, R. Goodwin, S. Murthy, R. Akkiraju, A. Kumaran and A. Derebail, "Scheduling Solutions for the Paper Industry," *Operations Research,* vol. 50, no. 2, pp. 249-259, 2002.

[5] D. Carlsson, S. D'Amours, A. Martel and M. Rönnqvist, "Supply Chain Planning Models in the Pulp and Paper Industry," *INFOR: Information Systems and Operational Research,* vol. 47, no. 3, pp. 167-183, 2009.

[6] F. Jaehn and R. Juopperi, "A Description of Supply Chain Planning Problems in the Paper Industry with Literature Review," *Asia-Pacific Journal of Operational Research,* vol. 36, no. 1, 2019.

[7] B. L. Golden and F. B. Alt, "Interval estimation of a global optimum for large combinatorial problems," *Naval Research Logistics Quarterly,* vol. 26, pp. 69-77, 1979.

[8] A. P. Giddings, R. R. Rardin and R. Uzsoy, "Statistical optimum estimation techniques for combinatorial optimization problems - a review and critique," *Journal of Heuristics,* vol. 20, pp. 329-358, 2014.

[9] K. Muralidhar and S. H. Zanakis, "A Simple Minimum-Bias Percentile Estimator of the Location Parameter for the Gamma, Weibull, and Log-Normal Distributions," *Decision Sciences,* vol. 23, pp. 862-879, 2007.